# A NEW FAMILY OF PLANETS?
# "OCEAN-PLANETS"

**A. Léger [1], F. Selsis [2], C. Sotin [3], T. Guillot [4], D. Despois [5],
H. Lammer [6], M. Ollivier [1], F. Brachet [1], A. Labèque [1], C. Valette [1]**

---------------

[1] Institut d'Astrophysique. Spatiale, bat 121, CNRS, Univ. Paris-Sud, F-91405 Orsay, Fr,
    tel: 33 1 69 85 85 80,    fax: 33 1 69 85 86 75
    Alain.Leger@ias.u-psud.fr, Marc.Ollivier@ias.u-psud.fr, Franck.Brachet@ias.u-psud.fr
    Alain.Labeque@ias.u-psud.fr, Claude.Valette@ias.u-psud.fr

[2] Centro de Astrobiología, 28850 Torrejón de Ardoz, Madrid, Sp.,
    tel: 34 91 520 64 30    fax: 34 91 520 16 21
    selsis@obs.u-bordeaux1.fr

[3] Géophysique, Université de Nantes, F-44321 Nantes cedex 3, Fr.,
    tel: 33 02 51 12 54 66    fax: 33 02 51 12 52 68
    Christophe.Sotin@chimie.univ-nantes.fr

[4] Observatoire de la Côte d'Azur, BP 4229, F-06304 Nice cedex 04, Fr.,
    tel: 33 4 92 00 30 47    fax: 33 4 92 00 31 21
    guillot@obs-nice.fr

[5] Observatoire de Bordeaux (INSU/CNRS), B.P. 89, F-33270 Floirac, Fr.,
    tel: 33 5 57 77 61 59    fax: 33 5 57 77 61 10
    despois@obs-bordeaux1.fr

[6] Space Res. Inst., Austrian Acad. of Sciences, Schmiedlstr. 6, 8042 Graz, Austria,
    tel:    0043 316 4120 641    fax:    0043 316 4120 690
    helmut.lammer@oeaw.ac.at

ABSTRACT

A new family of planets is considered which is in between rocky terrestrial planets and gaseous giant ones: "Ocean-Planets". We present the possible formation, composition and internal models of these putative planets, including that of their ocean, as well as their possible Exobiology interest. These planets should be detectable by planet detection missions as Eddington and Kepler, and possibly COROT (launch scheduled in 2006). They would be ideal targets for spectroscopic missions such as Darwin / TPF.

1. INTRODUCTION

The extrasolar planetary systems discovered thus far show a surprising diversity of orbital parameters. Most of these systems do not resemble our own Solar System. There is presently no consensus as to why these planetary systems are so diverse, but it seems that migration due to interactions between planets and the protoplanetary disk is an important ingredient (Lin et al., 1995; Ward, 1997; Trilling et al., 1998). It seems reasonable to assume that planets resembling our Uranus and Neptune, or slightly less massive ones, may have formed in cold regions of a protoplanetary disk and migrated inward, possibly into the so-called "Habitable-Zone" where liquid water can be present at their surface. These planets would be the more interesting as their large radius makes them rather easily detectable by Transit missions (COROT, Eddington, Kepler) and analysed by Darwin / TPF. A planet with twice the Earth radius requires, for spectroscopy analysis, an integration time 16 times shorter than an Earth analogue in the same distance and S/N conditions.

The interest of such planets is at least double, for Planetology and Exobiology. They would be a kind of objects we have not in the Solar System and significantly extend the field of Planetology. The search for a form of life similar to that which has developed on Earth would open a new field in Exobiology because the conditions of the environment would be quite different from the terrestrial ones. The minerals necessary to living bodies (P, S, Fe, Mg, Na, K…) could be brought to the surface by micro-meteorites or found in the ocean as dissolved species. It could even tell us something about origin of life on Earth e.g. by excluding an origin from black smokers at the silicate/ocean interface because on Ocean-Planets, liquid water and silicates would be separated by thousands of kilometres of ice.

It is relatively clear that around stars in which C/O is not too high, planetesimals built in the cold regions of the protoplanetary disk will contain a significant fraction of water ice. In our Solar System, this is the case of all the moons of the giant planets except Io. Uranus and Neptune can themselves be considered as "ice giants": their interior density is indeed very similar to that of compressed water ice (Podolak et al., 2002). However, Uranus and Neptune also contain about 1 to 4 Earth masses in hydrogen and helium in the form of an outer envelope. This implies that any primarily water layer that may exist is confined to the planet's deep regions, at temperatures (> 2000K) and pressures (> 10 GPa) too high to be even remotely reminiscent of an ocean. The case of the planets we consider in this Note is different as they contain much less hydrogen.

Of course, a lot of parameters drives the structure and composition of any planets. As planet formation is not well understood, and to limit the scope of this study, we choose to make the following assumptions:

1. only planets with mass in the range $1 < M / M_{Earth} < 8$ are considered. The lower boundary is for selection of planets that are easier to detected, and the upper one for objects that have not accreted a large amount of $H_2$ (Wuchterl et al., 2000). It is pointed out that the biggest objects of that type (M ~ 6-8 $M_{Earth}$) are of special interest because they are accessible to both Transit detection and Radial Velocity measurements so that *their radius and mass can be determined simultaneously*. Their composition is, in mass, 50% rocks and metals, 50% ices, as can be expected from the disc chemical composition;

2. the initial composition of ices is similar to that of comets, mostly $H_2O$ and some $NH_3$ and $CO_2$. To simplify the photochemistry, we assume no CO nor $CH_4$. The latter condition should be fulfilled if the planet forms further that the $H_2O$ ice boundary (T ~ 150 K) but closer than the ice boundary of the CO and $CH_4$, (T ~ 24 / 48 K and 30 / 58 K, respectively). The two values for the latter species correspond to the condensation of the pure material / its incorporation into clathrate hydrates (Iro et al., 2003). The composition is: m 90% $H_2O$, 5% $NH_3$ and 5% $CO_2$;

3. the planet is differentiated, and possesses, from the centre to the outside, a metallic core (mainly Fe + Ni), a rocky envelope (silicates), an ice/liquid layer and a gaseous atmosphere. Differentiation is likely due to the primordial heat acquired through impacts during the formation of the planet and gravity. Note that Io, Europa and Ganymede are differentiated, as all other major solid planets of the Solar System;

4. the interior of the planet is "hot", with an adiabatic temperature profile. Clearly, this is an assumption that needs further work. However, it represents a useful extreme baseline compatible with the relatively large masses and rapid inward migration of the object (less than ~10 Myr with type I migration; Ward, 1997);

5. the atmosphere has a given surface temperature governed by the distance to the star and the Greenhouse Effect. Its composition results from volatile outgasing, photo-chemistry and atmosphere escape. It is made of $N_2$, $H_2O$, and $CO_2$;

6. only planets that have migrated into the "Habitable Zone" are considered. The surface of the planet can be a huge ocean if its surface temperature is above the water triple point and under its critical one.

Now, questions are: (i) what is their internal structure? (ii) depth of ocean? (iii) atmospheric properties?

## 2. INTERNAL STRUCTURE

A model for planet interior is used that has been developed for the Earth interior and expanded to extrasolar terrestrial planets (Dubois, 2002). Quantities depend upon the radius, $r$, (1 D model). They are the planetary mass located between 0 and $r$, local density, gravity and pressure, $m(r)$, $d(r)$, $g(r)$, and $P(r)$, respectively.

Density is a function of material, pressure and temperature. The latter dependence is not major and is neglected. High-pressure laboratory experiments have been performed that provide equations of state $d(P)$. For the different materials, the adopted relations are as follows:

- main metals in the centre of a telluric planet are Fe and Ni. The equation of state of iron by Anderson & Ahrens (1994), is used;

- main kinds of silicates in the Solar System are Fe and Mg ones. Their phase at high pressure is Perovskite (P > 23 Mpa; Anderson, 1997) with the $d(P)$ relation as measured by Duffy and Arhens (1995);

- the phase diagram of $H_2O$ is complex, with 10 ice phases (Durham & Stern, 2001). The ice $d(P)$ relation has been measured up to 20 GPa by Fei et al. (1993) and we extrapolate it for higher pressures.

## 2.1 Internal Structure Modelling

Inputs are: the total mass of the planet, the fractions of metals, rocks and ices. Outputs are the outer planetary radius, the outer radius of the different material shells, variation with radius of the different physical quantities.

For a given mass of the planetary components, a self-consistent solution of the density, gravity and pressure functions, d($r$), g($r$), and P($r$), is required that fulfils the mass conditions

$$M_i = 4\pi \int_{R_{i-1}}^{R_i} d_i(r) \, r^2 \, dr \quad , \tag{1}$$

where the index, i, corresponds to the metal, rock and ice component, respectively, and $R_i$ the outer radius of component (i); the gravity and pressure relations are

$$g(r) = G \, m(r) / r^2 \tag{2}$$

$$P(r) = \int_r^{R_{pl}} d(x) \, g(x) \, dx \quad , \tag{3}$$

where $R_{pl}$ is the planetary radius. **For a 6 $M_{Earth}$ planet**, with the relative amounts of material as described in section 1 and deduced from the Earth's ones (Javoy, 1999) i.e. m 17% metals, 33% silicates and 50% ices, the internal structure calculated is shown in Fig. 1. **The planetary radius is $R_{pl}$ = 2.0 $R_{Earth}$**, central pressure 1 600 GPa and surface gravity 1.54 $g_{Earth}$.

(Figure 1)

## 3. OCEAN

When such a planet migrates into the Habitable Zone, its surface ice melts. A question appears, "how deep is the ocean?". As a first step, for the sake of simplicity an ocean of pure water is considered.

The temperature is a function of depth, $z$, with $z$ = $R_{pl}$ – $r$, and therefore pressure. The internal heat of the planet is transferred through the ocean. The **maximum temperature** *gradient* is adiabatic, otherwise convection in water would be trigged which is extremely efficient to transfer energy. We assume an adiabatic dependence for the temperature but it could be weaker. The adiabatic relation is:

$$dT/dP = \alpha T / (d_{water} C_p) \quad , \tag{4}$$

where $\alpha$ is the thermal expansion coefficient and $C_p$ the heat capacity. For P < 1-2 GPa, using the values $\alpha \approx 3\,10^{-4}$ K$^{-1}$, $d_{water} \approx 1$ g cm$^{-3}$, $C_p \approx 4$ J K$^{-1}$g$^{-1}$, the temperature reads:

$$T_{adiabat}(z) \approx T_{surface} + 20\, P(GPa) \tag{5}$$

The T[P(z)] position in the water phase diagram indicates whether it is liquid or solid, the P(z) relation depending upon the surface gravity, the liquid water zero pressure density, $d_o$, and compressibility, $\kappa$, P(z) = -$\kappa$ Ln(1 – g $d_o$ z / $\kappa$). The bottom of the ocean is obtained when the temperature curve crosses the solidus line. Its depth, bottom temperature and pressure result (Fig.2). For instance, assuming an adiabatic dependence of the temperature, a surface temperature, $T_{surf}$ = 7°C leads to an ocean depth of 72 km; the ocean bottom temperature and pressure are 35°C and 1.1 GPa. For a higher (lower) $T_{surf}$, the ocean depth would be larger (smaller), i.e. $T_{surf}$ = 30°C (0°C) leads to a 133 km (60 km) ocean depth.

Now, if the T dependence were less steep than adiabatic, the corresponding ocean would be more shallow. In the extreme case of a constant T (isothermal), its depth is 40, 45 and 65 km for $T_{surf}$ = 0, 7 and 30°C, respectively.

(Figure 2)

## 4. ATMOSPHERE

### 4.1 Composition

On Ocean-Planets, expected primary volatiles are $H_2O$, $NH_3$ and $CO_2$. The assumed initial composition of ice is similar to that of comets, e.g. m 90% $H_2O$, 5% $NH_3$ and 5% $CO_2$. Ammonia is very sensitive to UV, especially for 200 < $\lambda$ < 300 nm where the opacity due to other species is negligible. The fraction of the initial reservoir of $NH_3$ which happens to be in the atmosphere is photodissociated and converted into $N_2$ and $H_2$ in less than 2 Myr for a planet located at 1 AU of a G2V star. The produced hydrogen is subject to hydrodynamical escape and can sweep away a fraction of nitrogen.

Observations with the ASCA, ROSAT, EUVE, FUSE and IUE satellites of solar proxies indicate that a young GV star has continuous flare events producing a radiation environment several hundred times more intense than the solar one today (Guinan and Ribas, 2002). These results are in general agreement with former studies (Zahnle and Walker, 1992; Ayres 1997).

This high XUV flux is responsible for a large thermal escape rate of hydrogen resulting from photodissociation of water and ammonia. It is shown by Lammer et al. (2003a, b) that close-in exoplanets with hydrogen rich atmospheres develop an exosphere with a high temperature, $T_{exo}$, governed by heating by extreme UV, stellar particle flux and gas conduction, cooling by evaporation and expansion, the IR emission being negligible as the gas in such a region is optically extremely thin in IR. $T_{exo}$ obtained due to the XUV heating alone can be over 10 000 K which is much higher than the planet effective temperature, $T_{eff} = [L_* (1-A) / 4\pi \sigma_B D^2]^{1/4}$, where $L_*$ is the stellar luminosity, A the planetary albedo, $\sigma_B$ the Boltzmann constant and D the star-planet distance; e.g. for Earth, $T_{eff} \approx 255$ K.

The atmospheric escape is at least governed by the Jeans escape at $T_{exo}$. For temperatures in excess to 10 000 K, the kinetic energy of the atmospheric gas overcomes its gravitational one and the atmosphere expends and cools. Hydrodynamic conditions occur where the hydrogen flows off from the planet in bulk rather than merely evaporate from the exobase (Öpik,1963; Watson et al. 1981; Kasting and Pollack, 1983; Lammer et al. 2003a, b).

The 6 $M_{Earth}$ Ocean-Planet described above can release hydrogen into its atmosphere, via photodissociation of $NH_3$, with a maximum of 3 $M_{Earth}$ x 5 $10^{-2}$ x 3/17 = 1.6 x $10^{23}$ kg corresponding to an atmospheric pressure of ~ 1.2 GPa (1.2 $10^4$ bars). This upper value corresponds to the situation where the whole planetary ammonia is released into the atmosphere. Such a H amount can be blown off in a very short time (< 25 Myr) by energy limited hydrodynamic escape when the planet orbits at a distance of about 1 AU from a young G2V star (Guinan and Ribas, 2002; Lammer et al. 2003a).

After the majority of the hydrogen is lost and the heavier atmospheric compounds become relevant in the upper atmosphere, a part of the nitrogen should be dragged away by diffusion limited hydrogen flow (Zahnle et al. 1990). Because the amount of nitrogen left is not secure, the partial pressure of $N_2$ in the atmosphere is treated as a free parameter in the present paper. So is the $CO_2$ pressure.

### 4.2 Planetary albedo

Rayleigh back-scattering is efficient for a thick atmosphere e.g. with $P_{N2}$ = 10 bars. Fig. 3 shows the resulting minimum planetary albedo as calculated in a simple parallel plan model. Only a reduced fraction of the incoming stellar flux penetrates and heats the planet. This phenomenon *extends the Habitable Zone (HZ) to regions closer to the star*.

Kuchner (2003) recently calculated an additional protective process. If the atmosphere is still thicker and contains a huge amount of water vapour, it becomes absorbing in the visual-near IR and the heat transfer from its upper part to the (liquid) ground is slowed down. According to the author, this prevents the planet from transforming into a gas ball for billions of years, even if it is pretty close to its star.

### 4.3 Planet IR emission

The planet IR emission must be calculated in a self-consistent way as a function of the atmospheric composition and the stellar irradiation. A key ingredient is the temperature profile. It is reminded that an isothermal atmosphere would have no spectral feature, whatever its composition. As a preliminary calculation, for a given ground temperature, an atmosphere saturated with water vapour is considered and an adiabatic decrease of T is calculated down to an arbitrary value of 180 K. This limitation is to prevent T from reaching unrealistic low values.

Fig. 3 shows the interesting case of an Ocean-Planet with a 320 K (47°C) surface temperature. This case of a rather hot planet with a super-humid atmosphere, has no analogue in the Solar System. Its cold tropospheric trap (T ~ -65°C) is at 15 km. $H_2O$ absorption is very strong: at the hot low altitudes water bands are widened by pressure, and at high altitudes it keeps on absorbing as its abundance is still high.

(Figure 3)

### 4.4 <u>Biosignatures?</u>

As in Earth-like planets, photodissociation of $H_2O$ can produce abiotic $O_2$. This gas can accumulate, the more as the removing processes present on Earth, oxidation of rocks by soil weathering and oxidation of volcanic gases, are not expected on an Ocean-Planet. Consequently, ***the presence of $O_2$ is not a reliable biosignature.***

Can $O_3$ provide a better one? If $O_2$ is produced thanks to $H_2O$ photolysis at high altitude, hydrogenous compounds like H, OH and $HO_2$ are produced which attack very efficiently $O_3$ and prevent its accumulation. The only way to have a significant amount of $O_3$ in the atmosphere spectrum is that $O_2$ is produced at low altitude, e.g. by biological photosynthesis, and that no $H_2O$ gets at high altitude where UV are present. Consequently, as on terrestrial planets (Selsis et al., 2002), the simultaneous presence of $O_3$, $H_2O$ and $CO_2$ in the atmosphere appears to be a reliable biosignature. ***This points out the superiority of $O_3$ as a biosignature with respect to $O_2$.***

## 5. CONCLUSION

We have shown that massive ice-rich planets possibly form in external regions of protoplanetary disks and migrate inward. Depending on their distance to the star and properties of their atmospheres, some of them may form a surface water ocean. Such an ocean of liquid water would have a thickness of ~100 km.

Mid-future space missions searching for planetary transits in the Habitable Zone (Eddington, Kepler) coupled with Radial Velocity follow-up should provide us with valuable information about their existence and properties. If there are as resistant with respect to evaporation and photolysis of their atmospheres as some models predict, COROT (launch scheduled in 2006) should detect the hottest ones.

**ACKNOWLEDGMENTS**
We are grateful to Cécile Engrand and Olivier Grasset for their valuable help during this work

**Figure captions**

Figure 1: calculated internal structure of a 6 $M_{Earth}$ Ocean-Planet. Constituents are, from the centre to the outside, 1 $M_{Earth}$ metals, 2 $M_{Earth}$ rocks and 3 $M_{Earth}$ ice. The upper layer is an ocean (sect. 3). For comparison, the structure of the Earth calculated with the same model is shown, it closely fits the actual one. In the former planet, with density expressed in g cm$^{-3}$ and pressure in GPa, the ice density varies from 1.5 (P = 1) to 3.9 (P = 250), the silicate from 6.2 (P = 250) to 8.2 (P = 750), and the metal from 15.6 (P = 750) to 19.5 (P = 1 600).

Figure 2: Water temperature as a function of pressure. The full lines describe the phase diagram, and the dotted one the temperature adiabatic profile in the ocean that results from a given surface value (280 K = 7°C). As explain in the text, this profile is the steepest one expected, the actual one could be flatter. Within these conditions, the ocean depth is 72 km, the ocean bottom temperature and pressure are 35°C and 1.1 GPa, respectively. For a higher (lower) surface temperature, the ocean depth would be larger (smaller). A flatter T profile would lead to a more shallow ocean.

Figure 3: light reflected and emitted by Ocean-Planets versus wavelength. In the UV-visible, the minimum fraction of the stellar flux back-scattered by the planet is shown, for different values of the ground pressure of $N_2$, this gas being the main atmospheric component. That fraction is minimum because a zero surface albedo is assumed. If the latter is finite, the stellar flux back-scattered is higher. Different $N_2$ pressures are considered. In the IR, the emission is from a 6 $M_{Earth}$ planet with $T_{surface}$ = 320 K = 47°C, $P_{N2}$ = 5 bars and an atmosphere saturated with water vapour. It contains $CO_2$ with a mixing ratio of $10^{-5}$. As it is wet, and pretty warm at its surface, the planet is assumed to be cloudy with a full coverage of clouds at h = 5 km. Note the $CO_2$ absorption band at 15 μm and the strong $H_2O$ ones at λ < 8 μm and λ > 16 μm.

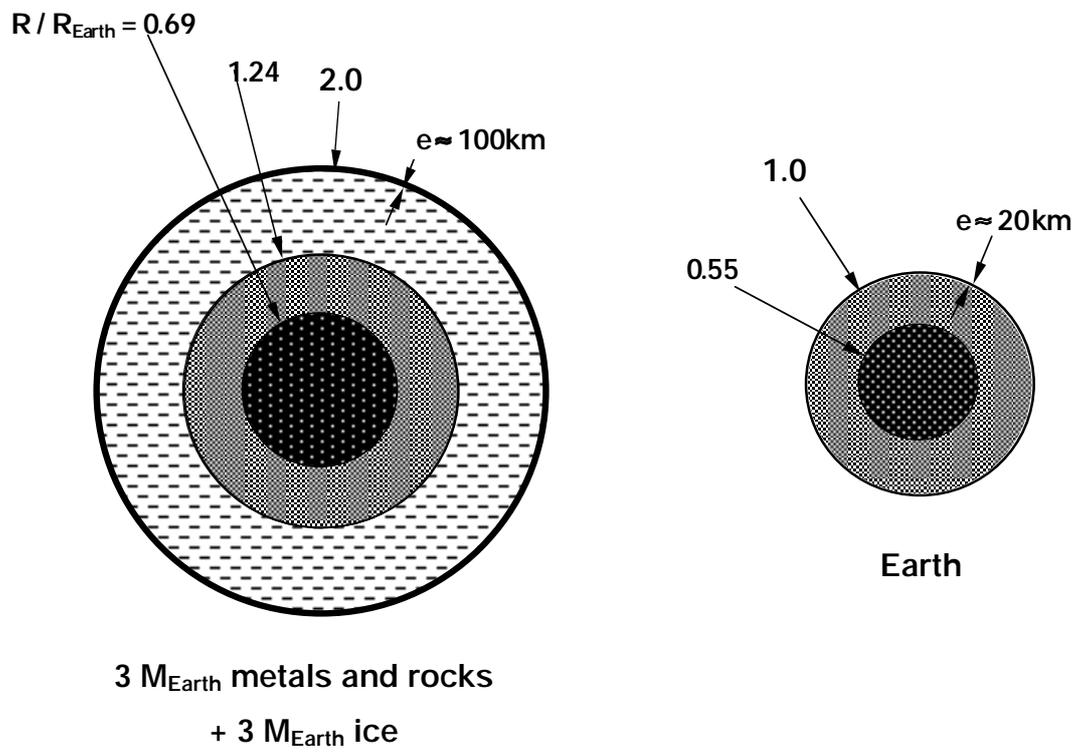

Figure 1

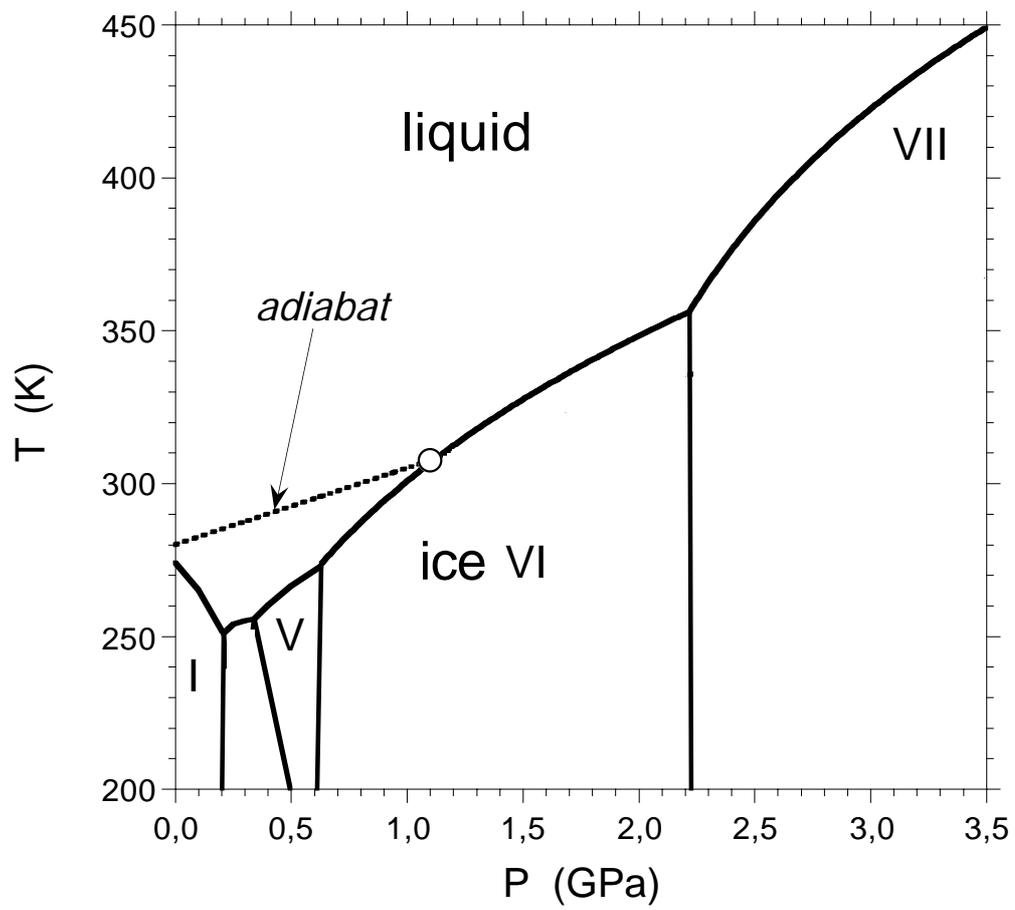

Figure 2

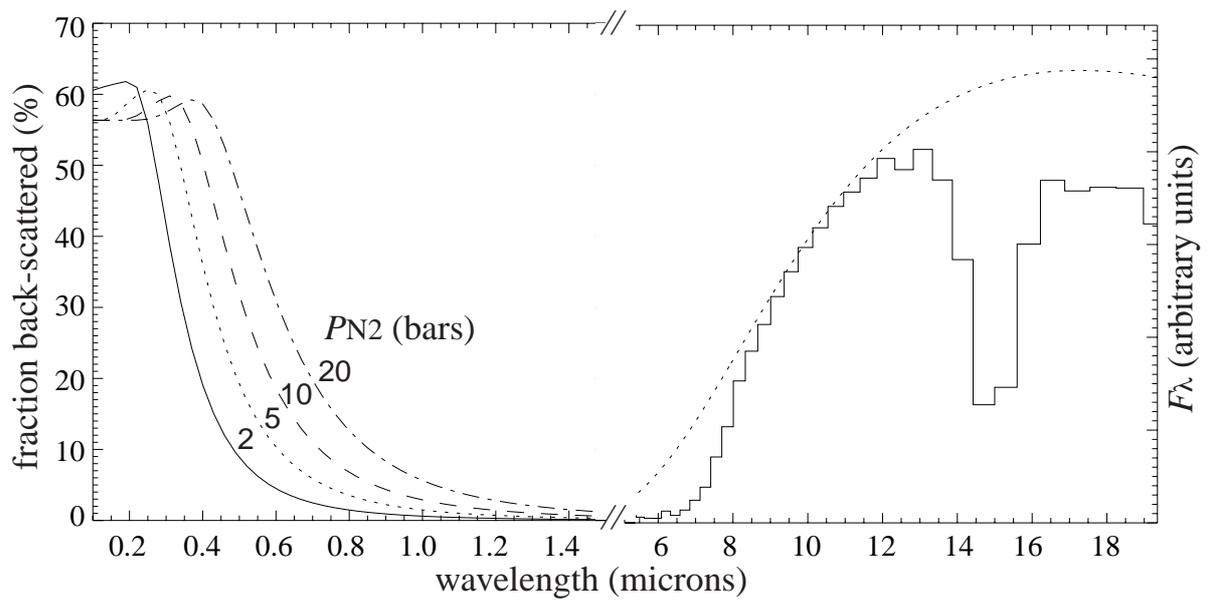

Figure 3